\documentclass[12pt,twoside]{article}
\usepackage{a4}\usepackage{epsf}\usepackage{latexsym}
\usepackage{cyrillic}\usepackage{Abbreviations}
\begin{document}
\title{ Phase transition in  scalar $\phi^{4}$-theory
 beyond the super daisy resummations } 
\author{
{\sc M. Bordag}\thanks{e-mail: Michael.Bordag@itp.uni-leipzig.de} \\
\small  University of Leipzig, Institute for Theoretical Physics\\
\small  Augustusplatz 10/11, 04109 Leipzig, Germany\\
\small and\\
{\sc V. Skalozub}\thanks{e-mail: Skalozub@ff.dsu.dp.ua}\\
\small  Dniepropetrovsk State University, 49050 Dniepropetrovsk, Ukraine}
\maketitle
\begin{abstract}
  The temperature phase transition in scalar $\phi^4(x)$ field theory with
  spontaneous symmetry breaking is investigated in a partly resummed
  perturbative approach. The second Legendre transform is used and the
  resulting gap equation is considered in the extrema of the free energy
  functional.  It is found that the phase transition is of first order in the
  super daisy as well as in a certain beyond super daisy resummations. There
  no unwanted imaginary parts in the free energy are found but a loss of the
  smallness of the effective expansion parameter near the phase transition
  temperature is found in both cases. This means an insufficiency of the
  resummations or a deficit of the perturbative approach.
\end{abstract}
\thispagestyle{empty}
\section{Introduction}\label{Sec1}
The question on the order of the phase transition when a spontaneously broken
symmetry is restored by raising temperature is a topic of considerable
interest.  The present status of the problem in scalar $\phi^{4}$-theory with
spontaneous symmetry breaking due to the mass term with 'wrong sign' is
characterized by the general opinion that it is a second order transition, see
e.g. the standard textbooks \cite{zinnjustin,linde1,Kapusta}. This opinion is
based on non-perturbative methods like lattice calculations and the average
action resp. flow equation method \cite{Tetradis:1993xd,Reuter:1993rm},
however still lacks a formal, final proof.

In the perturbative approach numerous authors found a first order phase
transition. Here the problem is in the infrared divergencies appearing
there. A resummation of the relevant diagrams is necessary. This was known
already since the pioneering papers in the field, e.g.,
\cite{DolanJackiw}. The general outcome is that the perturbative methods do
not give reliable results near the phase transition for two reasons:
\begin{enumerate}
\item Unwanted imaginary parts, for instance in the nontrivial minimum of the
  free energy (resp. the effective potential) appear indicating a instability
  or insufficient resummation. So for example, in the paper
  \cite{Espinosa:1992gq} a first order transition had been found on the level
  of   super daisy graphs resummed. But then the authors included higher loop
  graphs and found an unwanted imaginary part. 
\item Near the phase transition, the effective expansion parameter of the
  resummed perturbative series may become of order one making the
  contributions of higher order graphs unpredictable. For instance, it cannot
  be excluded that their inclusion may change the order of the phase
  transition.
\end{enumerate}
  
In the present paper we consider the scalar $\phi^{4}$-theory in (3+1)
dimensions in the perturbative approach. We perform a resummation of the
perturbative series using the second Legendre transform and go beyond the
super daisy resummation.  We refine our previous results
\cite{bordagskalozub99} obtained by a simpler resummation method which is
restricted to the level of including the super daisy diagrams only. The
disappearance of all unwanted imaginary parts was shown there and, thus, is
repeated in Sec.  \ref{sec3}. We note that this result implies that the
imaginary part found in \cite{Espinosa:1992gq} from higher loop graphs is
questionable.

Below, in Sec. \ref{sec4}, we calculate a number of contributions beyond the
super daisy level. We show that, near the phase transition, they are, as
expected, of the same order as the leading contributions. They turn out not to
change the order of the phase transition but to affect its quantitative
characteristics only making the transition a bit stronger first order.

To explain the controversy results on the order of the phase transition one
could argue that the higher loop graphs not included in the perturbative
approach might sum up to contributions that weaken the first order character
until it turns into a second order one when all graphs are summed up finally.
Our result, however, points in the opposite direction.

As technical tools we use the imaginary time formalism and functional methods.
The most important tool is the use of the second Legendre
transform\footnote{Is known since the beginning of the sixties
  (\cite{Luttinger60,Baym62} and can be found in textbooks \cite{vasiliev},
  see also \cite{2PI}).} together with the necessary condition for an extremum
of the free energy.

We consider the scalar $\phi^{4}$-theory in (3+1) dimensions with the 'wrong
sign' in the mass term which takes after symmetry breaking by shifting the
field $\phi(x)\to { v}+\phi(x)$ the form
\bea\label{Sone}
S[\phi]&=&\frac{m^{2}}{2}v^{2}-\frac{\la}{4}v^{4} \\
&&+\int \d x \left(\frac12
  \phi(x){\bf K}_{\mu}\phi(x)-\la v
  \phi(x)^{3}-\frac{\la}{4}\phi(x)^{4}+v(m^{2}-\la v^{2})\phi(x)\right) \nn
\eea
with ${\bf K}_{\mu}=\Box -\mu^{2}$ and the tree mass $\mu^{2}=-m^{2}+3\la
v^{2}$. 
 The line $\Delta$ (free propagator) in the corresponding
graphs is the inverse of $\bf K$: $\Delta= -{\bf K}^{-1}$. The quantity to be
calculated is the free energy $F$. By means of $W=-TF$ it is related to the
 vacuum Green functions 
\be\label{Z}
Z=\int D\phi ~{e^{S}}\,,
\ee
where $T$ is the temperature and  its connected part reads
\be\label{W}
W=\log Z \,.
\ee
After performing the second Legendre transform the representation   
\be\label{Wallg}
W=S[0]+\frac12 Tr \log \beta -\frac12 Tr \Delta^{-1}\beta +W_{2}[\beta]
\ee
emerges, where all graphs are to be taken with the line $\beta$
(instead of $\Delta$) which is subject to the Schwinger-Dyson (SD) equation
\be\label{SDyallg}
\beta^{-1}(p)=\Delta^{-1}-\Sigma[\beta](p) \,,
\ee
where
\be\label{Sigallg}
\Sigma[\beta](p)=2{\delta W_{2}\over \delta \beta(p)}
\ee
are the self energy graphs with no propagator insertions, see, e.g., 
\cite{2PI} for details. $W_{2}[\beta]$ is the sum of all two particle
irreducible (2PI) graphs . 

In the imaginary time formalism  the operation  '$Tr$' is given by
\be\label{Tr}
Tr_p=T\sum_{l=-\infty}^{\infty}\int {\d^{3}\vec{p}\over (2\pi)^{3}}
\ee
with the momentum is $p=(2\pi T l, \vec{p})$.  We indicate the dependence on a
functional argument $\beta$ by square brackets and on the momentum $p$ by
round ones ($\beta(p)$ is the corresponding Fourier transform), $\delta$ is
the variational derivative.

Turning to perturbation theory it is useful to start from $W_{2}$.  The first
contribution to its perturbative expansion is the graph looking like a
'eight',
\be\label{W2-1}
W_{2}^{(1)}=\frac18 \epsfxsize=0.5cm\raisebox{-8pt}{\epsffile{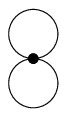}}\,.
\ee
By means of the formulas given above it generates the free energy with all
super daisy diagrams summed up. In this way the results of our previous paper
\cite{bordagskalozub99} are reproduced.  In the present paper we go beyond the
super daisy level by including the following (higher loop) graphs
\be\label{W2-h}
W_{2}^{(h)}=
\frac18 \ \epsfxsize=0.5cm\raisebox{-8pt}{\epsffile{eight.eps}}
+\frac{1}{12}\ \epsfxsize=0.5cm\raisebox{-4pt}{\epsffile{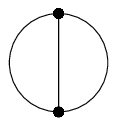}}
+\frac{1}{48}\ \epsfxsize=0.5cm\raisebox{-4pt}{\epsffile{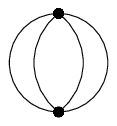}}
+\sum_{n\ge 3}{1\over  2^{n}}\ \epsfxsize=0.5cm\raisebox{-4pt}{\epsffile{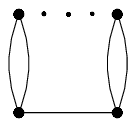}}
+\sum_{n\ge 3}{1\over 2^{n+1}n}\
\epsfxsize=0.5cm\raisebox{-4pt}{\epsffile{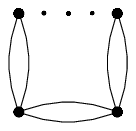}} \,,
\ee
where the dots symbolize further subgraphs to be inserted so that the number
of vertices is $n$. This selection of graphs is motivated by the 1/N expansion
where it includes all leading and first non-leading contributions.
\section{The SD equation and the extrema of the free energy}\label{sec2}
The problem to be considered with respect to the phase transition is whether
the free energy exhibits a nontrivial (i.e., at $v\ne 0$) maximum
\underline{and} minimum for some $v>0$ at finite $T$.  The common approach is
to solve the SD equation \Ref{SDyallg} in some approximation and to insert its
solution into the free energy \Ref{Wallg} with a subsequent investigation of
its dependence on the condensate $v$. Instead, we suggest to consider the
necessary condition for the extremum of the free energy
\be\label{cond1}
{\pa W\over \pa v^{2}}=0
\ee
together with the Schwinger-Dyson equation \Ref{SDyallg}.  In this way one
variable can be eliminated and we obtain a simpler expression. In fact there
are more simplifications behind. Away from the extrema of the free energy
there are tadpole graphs generated by the linear term in the action \Ref{Sone}
which must be taken into account. They disappear once Eq. \Ref{cond1} holds.

We obtain from Eq. \Ref{cond1} using \Ref{Wallg} and  $\Delta^{-1}=-{\bf
K}_{\mu}^{-1}=p^2-m^2+3\la v^{2}$ the condition
\be\label{1}
{\pa W\over \pa v^{2}}={m^{2}\over 2}-{\la\over 2}v^{2}-\frac32 \la Tr \beta
+Tr {\pa \beta\over \pa v^{2}}\left(\frac12 \beta^{-1}-\frac12 {\bf K_{\mu}}+{\delta
    W_{2}\over \delta \beta}\right)+{\pa W_{2}\over \pa v^{2}} = 0\,,
\ee
which by means of \Ref{SDyallg} reduces\footnote{This is related to the known
  fact that the SD equation \Ref{SDyallg} is a stationary point of $W$
  considered as a functional of $\beta$, see e.g. \cite{Blaizot:1999ip}.} to
\be\label{lav}
\la v^{2}= m^{2}-3\la \Sigma_{0}+2{\pa W_{2}\over \pa v^{2}} 
\ee
with the notation
\be\label{Sig0}
\Sigma_{0}=Tr \beta \,.
\ee
A representation of $\Sigma_{0}$ is given in the Appendix.

Now the SD equation \Ref{SDyallg} can be rewritten eliminating $v$ using
\Ref{lav} to become
\be\label{SDyalgg1}
\beta^{-1}=p^{2}+2m^{2}-9\la\Sigma_{0}+6{\pa W_{2}\over \pa
  v^{2}}-\Sigma[\beta](p) \,.
\ee
It holds in the extrema of the free energy and allows directly to determine
number and kind of the extrema. 

Being interested in the infrared behavior at high temperature we make the
ansatz 
\be\label{Ansatz}
\beta^{-1}(p)=p^{2}+M^{2}
\ee
and obtain from   Eq. \Ref{SDyalgg1} the gap equation in the extremum
\be\label{gapone}
M^{2}=2m^{2}-9\la\Sigma_{0}+6 {\pa W_{2}\over \pa
  v^{2}}-\Sigma[\beta](0)\,,
\ee
where the ansatz \Ref{Ansatz} is assumed to be used for $\beta$ in the
right hand side.

To proceed we need now to make the approximation defined by Eq. \Ref{W2-h} on
the level of the 2PI functional.  The analytic expression for $W_{2}^{(h)}$
corresponding to Eq. \Ref{W2-h} can be obtained from the graphical
representation by inserting the line $\beta$ and the vertex factors $-6\la v$
resp.  $-6\la$ for the three resp. four vertex. It reads
\be\label{W2-bc}
W_{2}^{(h)}=-\frac34 \la \Sigma_{0}^{2}+3\la^{2}v^{2}\Gamma
+\la^{2 }D \,,
\ee
where we introduced the notations
\be\label{Ga}
\Gamma=Tr_{q} \beta(q)\Sigma_{1}(q){1-6\la\Sigma_{1}(q)\over 1+3\la\Sigma_{1}(q)}
\ee
and
\be\label{D}
D=\frac34Tr_{q} \Sigma_{1}(q)^{2}\left(1+6\sum_{n\ge
    1}{(-3\la\Sigma_{1}(q))^{n}\over n+2 }\right)
\ee
as well as
\be\label{Sig1} \Sigma_{1}(p)=Tr_{q}
\beta(q)\beta(p+q)= ~~ \epsfxsize=0.9cm\raisebox{-3pt}{\epsffile{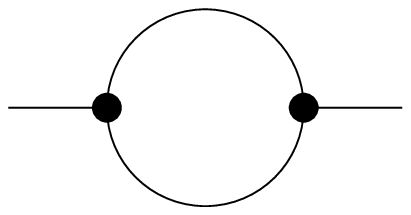}} ~~~.
\ee
Note that the infinite set of diagrams in Eq. \Ref{W2-h} can be reproduced by
a formal expansion of \Ref{W2-bc} with respect to powers of $\la$.

From these formulas we obtain by means of \Ref{Sigallg} the self energy in the
form
\be\label{Sigone}
\Sigma[\beta](p)=-3\la\Sigma_{0}+6\la^{2}v^{2}{\delta \Gamma\over
  \beta(p)}+2\la^{2}{\delta D\over \delta \beta(p)},
\ee
where we used obvious relations like
$\delta\beta(p)/\delta\beta(q)=\delta(p-q)$ and
$\delta\Sigma_1[\beta](p)/\delta\beta(q)=2\beta(p+q)$. 
With ${\pa W_{2}^{(h)}\over \pa v^{2}} =3\la^{2}\Gamma$ and Eq. \Ref{Sigallg}
we obtain the SD equation in the extremum
\be\label{SDyone}
\beta^{-1}=p^{2} +2\left(m^{2}-3\la\Sigma_{0}+6 \la^{2}\Gamma\right)
\left(1-3\la{\delta\Gamma\over\delta \beta(p)}\right)+6\la^{2}\Gamma -
2\la^{2}{\delta D\over\delta \beta(p)} \,,
\ee
and from \Ref{gapone} the gap equation
\be\label{gaptwo}
M^{2}=2\tilde{\Gamma} \left( m^{2}-3\la\Sigma_{0}+6\la^{2} \Gamma \right)\,
\ee
in the given approximation.
Here we used the relation 
\be\label{relp}
6\la^{2}\Gamma -
2\la^{2}{\delta D\over\delta \beta(p)}=6\la^{2}Tr_{q} \left(\beta(q)-\beta(p+q)\right)\Sigma_{1}(q){1-6\la\Sigma_{1}(q)\over 1+3\la\Sigma_{1}(q)}
\ee
showing that this contribution to Eq. \Ref{SDyone} vanishes for $p=0$. The
quantity $\tilde{\Gamma}$ is given by
\be\label{Gatilde}
\tilde{\Gamma}\equiv1-3\la{\delta\Gamma\over\delta \beta(0)}
=
{1-6\la\Sigma_{1}(0)\over 1+3\la\Sigma_{1}(0)}
+18\la Tr_{q} \beta(q)^{2}{ 3\la\Sigma_{1}(q)(2+3\la\Sigma_{1}(q))\over
(1+3\la\Sigma_{1}(q))^{2}} \,,
\ee
where $\Gamma $ \Ref{Ga} was used. 

Note that Eq. \Ref{gaptwo} does not contain the condensate $v$ so that it is
an equation to determine the mass $M$ in the extrema of the free energy as a
function of $\la$, $T$ and $m$. 

\section{Symmetry restoration  in super daisy approximation}\label{sec3}
%
Now, in order to establish the connection to our previous results
\cite{bordagskalozub99} and to make the subsequent calculations more
transparent we restrict ourselves for a moment to the level of super daisy
graphs summed up only. This is obtained by setting $\Gamma$ and $D$ in the
formula of the preceding section equal to zero. So, with
\be\label{W1} W_{2}^{(1)}=\frac18 \ 
\epsfxsize=0.5cm\raisebox{-8pt}{\epsffile{eight.eps}}=-\frac34\la
\Sigma_{0}^{2} \ee
we obtain the gap equation in the extremum in the form 
\be\label{gapeqsd}
M^{2}=2m^{2}-6\la\Sigma_{0}
\ee
(cf. Eq. (37) in \cite{bordagskalozub99}).  For the propagator
$\beta^{-1}=p^{2}+M^{2}$ is exact because $\Sigma_{0}$ does not depend on the
momentum $p$.  Equation \Ref{lav} for the condensate reduces to
\be\label{lav1}
\la v^2 = m^{2}-3\la \Sigma_{0}
\ee
so that we can note
\be\label{lav2}\la v^2 ={M^2\over 2}
\ee
in the given approximation.

The solution of equation \Ref{gapeqsd} can be investigated graphically by drawing both
sides of the equation as function of $M$ in one plot. A detailed analysis had
been given in \cite{bordagskalozub99} showing clearly the behavior expected for a first order
phase transition. For instance, in a certain temperature region the existence
of two solutions corresponding to a maximum and a minimum of the free energy
was found.
For small coupling $\la$, which corresponds to high temperatures of the phase
transition $T\sim\la^{-1/2}$, the function $\Sigma_{0}$ in the gap equation can
be expanded in powers of $T/M$, see Eq. \Ref{d0a} in the appendix for details. Keeping the
relevant orders the gap equation turns into   
\be\label{Mquadr}
M^{2}=2m^{2}-6\la\left({T^{2}\over  12}-{MT\over 4\pi} +{m^{2}\over
    8\pi^{2}}+{M^{2}\over 16\pi^{2}}\left(\log{(4\pi T)^{2}\over 2m^{2}}-2\gamma\right)\right)
\ee
and has the solutions
\be\label{M+-}
M_{_{{Min\atop Max}}}={3\la T\over 4\pi \delta_{3}}\pm
\sqrt{{1\over
  \delta_{3}}}
\sqrt{
\left({3\la T\over 4\pi }\right)^{2}{1\over
  \delta_{3}}+{2m^{2}\delta_{1}}-{\la T^{2}\over 2} } \,,
\ee
where the notations
\be\label{d1}
\delta_{1}=1-{3\la\over 8\pi^{2}}
\ee
and
\be\label{d3}
\delta_{3}=1+{3\la\over 8\pi^{2}}\left(\log{(4\pi T)^{2}\over
    2m^{2}}-2\gamma\right) \,.
\ee
are used.  Here, $M_{_{Max}}$ resp. $M_{_{Min}}$ (and $v$ by means of Eq.
\Ref{lav2}) correspond to the minimum resp. to the maximum of the free energy
at $v\ne 0$. The lower spinodal temperature $T_{-}$, i.e., the temperature at
which the maximum appears at $v=0$, follows from $M_{Max}=0$ to be
$T_{-}={2m\over\sqrt{\la}}\sqrt{\delta_{1}}$ and the upper spinodal
temperature $T_{+}$, i.e., the temperature where the minimum and the maximum
merge and disappear for $T>T_{+}$ follows from \Ref{M+-} to be $T_{+}=
{2m\over\sqrt{\la}}\sqrt{{\delta_{1}\over 1-{9\la\over 8\pi^{2}\delta_{3}}}}$.
For small $\la$, in the first nontrivial order, they are
\be\label{T+-}
 T_{-}={2m\over\sqrt{\la}}\left(1-{3\la\over 16\pi^{2}}\right),
\qquad T_{+}={2m\over\sqrt{\la}}\left(1+2{3\la\over 16\pi^{2}}\right)\, ,
\ee
and their ratio 
\be\label{ratio}
{T_{+}\over T_{-}}=1+{9\la\over 16\pi^{2}}\ee
is larger than one. We note that this does not depend on the logarithm of $T$
coming in with $\delta_3$. From Eq. \Ref{M+-} it is easy to calculate the
masses in $T_{-}$ and in $T_{+}$:
\bea\label{Mia+-}
{M_{_{Min}}}_{|_{T=T_{-}}}&=&{3\sqrt{\la}\over\pi}m, \qquad
{M_{_{Max}}}_{|_{T=T_{-}}}=0 \,, \nn \\
{M_{_{Min}}}_{|_{T=T_{+}}}&=&{M_{_{Max}}}_{|_{T=T_{+}}}=
{3\sqrt{\la}\over 2 \pi}m \,.
\eea
The temperature $T_{c}$ of the phase transition where the free energy is the
same in the trivial (i.e., at
$v=0$) and in the nontrivial minimum can be calculated in the given
approximation quite easy. For this  one needs the gap equation
\Ref{SDyallg} at
$v=0$, $\beta^{-1}=p^2-m^2+3\la\Sigma_{0}$. Note, that the self energy in
the given approximation   
\be\label{S1}
\Sigma(p)=2{\delta\over\delta \beta(p)}\left(-\frac34\la\Sigma_{0}^{2}\right)=
-3\la\Sigma_{0} 
\ee
is independent of $p$ and the ansatz $\beta^{-1}(p)=p^{2}+M_{0}^{2}$
is exact again, where $M_{0}$ is the mass as solution of the gap
equation at $v=0$. In this way the gap equation there reads
$M_{0}=-m^{2}+3\la\Sigma_{0}$. Taking the approximation \Ref{d0a} for
$\Sigma_{0}$ we obtain in a way similar to Eq.  \Ref{M+-} its
solution: $M_{0}={-3\la T\over 8\pi \delta_{3}}+ \sqrt{ \left({3\la
T\over 8\pi \delta_{3}}\right)^{2}-{m^{2}\delta_{1}\over
\delta_{3}}-{\la T^{2}\over 4\delta_{3}} }$.

Now $T_{c}$ can be obtained from equating the free energy at $v=0$ and in the
minimum:
\be\label{condTc}
W_{0}^{(1)}=W_{e}^{(1)} \quad \mbox{with} \quad  \left\{ {
\begin{array}{rcl}
W_{0}&=&W(v\to 0,M\to M_{0})\\
 W_{e}&=&W(v\to \sqrt{M_{Min}/2\la},M\to M_{Min})\,,
\end{array}} \right. \,
\ee
where $W$ are the connected Green functions \Ref{Wallg}. We rewrite them using
$\Delta^{-1}\beta=1-\beta\Sigma$ following from the Schwinger-Dyson equation
\Ref{SDyallg} and drop the infinite constant $Tr 1$ appearing thereby. Again,
in the given approximation, \Ref{W1}, using \Ref{S1} we obtain
\be\label{W1a}
W^{(1)}=-{m^{2}\over 2}v^{2}+{\la\over
  4}v^{4}-\frac34\la\Sigma_{0}^{2}+\frac12 \Tr\ln\beta \,. 
\ee

Now, $W_{0}^{(1)}$ resp. $W_{e}^{(1)}$ are obtained by inserting the
corresponding values of $v$ and $M$. After that, Eq. \Ref{condTc} can be
approximated in the lowest nontrivial order in $\la$ (thereby, in $ \frac12
\Tr\ln\beta$ all orders of $\la$ shown in \Ref{V1a} have to be taken). The
temperature $T_{c}$ is in between $T_{-}$ and $T_{+}$, closer to $T_{+}$. With
$T_{c}=T_{+}(1-\la \tau)$ we obtain $\tau=0.0028$ as numerical solution of the
emerging higher order algebraic equation.

So, on this stage of resummation, i.e., after summation of all super
daisy diagrams which is generated by taking for $W_{2}$ the the representation
$W^{(1)}$ \Ref{W1}, we see clearly a first order phase transition. In the
nontrivial minimum the mass $M_{Min}$ \Ref{M+-} is positive and of order
$\sqrt{\la}$:
\be\label{rangeM}
{3\over 2\pi}\sqrt{\la}m ={M_{Min}}_{|_{T_{-}}} \le M_{Min}\le
{3\over \pi}\sqrt{\la}m ={M_{Min}}_{|_{T_{+}}}\,.
\ee
This mass has to be inserted into all higher loop graphs which are not
included in the given approximation. As the theory is Euclidean and the
propagator is $\beta(p)=1/(p^{2}+M^{2})$ a consequence is the absence of
unwanted imaginary parts.  In this way the first problem mentioned in
the introduction is settled.

Consider now the second. A simple counting of the expansion parameter of the
perturbative series after all super daisy diagrams are summed up goes as
follows. Let $V$, $C$ resp. $L$ be the number of vertices, loops resp. line of
a graph. The vertex factors are proportional to $\la$ ($\phi^{4}$-vertices)
resp. $\la v$ ($\phi^{3}$-vertices). By means of formulas \Ref{lav2} and
\Ref{Mia+-} the condensate is of order one, $v\sim m$, so that all vertex
factors are of order $\la$. The summation/integration measure associated each
loop is $T\sum_{l=-\infty}^{\infty}\int \d \vec{p}/(2\pi)^3$ \Ref{Tr} and the
factors of the lines are $1/((2\pi l T)^{2}+\vec{p}^{2}+M^{2})$. For high $T$
($\sim\la^{-1/2}$) \Ref{T+-} the graphs can be approximated by taking the
zeroth Matsubara frequency ($l=0$).  Than, doing a rescaling $\vec{p}\to
\vec{p}M$, a factor $A=\la^{V}T^{C}M^{3C-2L}$ appears in front of each graph
which itself is then a number of order one. By means of $C=L-V+1$ and $M\sim
\sqrt{\la}$ we obtain $A\sim\la$ so that it ceases to depend on the order of
the perturbative expansion. This is a loss of the smallness of the effective
expansion parameter. Although we know that all graphs are finite and of order
one, we do not have a control over the sum of them. For instance, it cannot be
excluded that summing up further (or all) graphs the order of the phase
transition may change. Thus, super daisy approximation is insufficient to solve
this problem.

\section{Beyond the  super daisy resummation}\label{sec4}
%
To account for graphs beyond the super daisy approximation let us
return to the Schwinger-Dyson equation in the extremum
\Ref{SDyone}. Now, because in the r.h.s. there are contributions
depending on $p$, the simple ansatz \Ref{Ansatz} is strictly speaking
insufficient. However, being interested in the infrared behavior it is
anyway meaningful to use it at least in order to get a qualitatively
correct result. After that we arrive at the gap equation
\Ref{gaptwo}. Here, $\Gamma$ \Ref{Ga} and $\tilde{\Gamma}$
\Ref{Gatilde} contain the the higher loop graphs. In order to get an
estimate of these quantities at high $T$ we make the following
approximations.

Consider first $\Gamma$ at $\la=0$. It is the simple 2-loop graph
$\Gamma(\la=0)= \epsfxsize=0.5cm\raisebox{-4pt}{\epsffile{orech.eps}}$. Taking
the contribution resulting from the zeroth Matsubara frequency only, it
becomes $\Gamma(\la=0)\sim \gamma T^{2}$ with some   $\gamma$ which we do not
calculate explicitely. In fact, $\gamma$ may exhibit a logarithmic dependence
on $T$ which we ignore, however. After that we observe that $\Sigma_{1}(q)$,
for large $q$, behaves logarithmically (its superficial divergence degree is
zero). Thereafter we approximate it by the first term of its hight $T$
expansion \Ref{sig1a}. It is a constant. Hence,  we obtain
for $\Gamma$ \Ref{Ga} in this approximation 
\be\label{gaapprox} \Gamma ={1-2\ep\over 1+\ep} {Tr_{q}} \beta(q)\Sigma_{1}(q)
\sim {1-2\ep\over 1+\ep} \gamma T^{2}\,, 
\ee
where we used \Ref{sig1a} and introduced the notation $\ep \equiv
  {3\la T\over 8\pi M}$ so that $3\la\Sigma_{1}(0)\sim \ep$ holds.

By the same procedure we obtain for $\tilde{\Gamma}$ \Ref{Gatilde} the
approximation
\be\label{gatapprox}
\tilde{\Gamma}(\ep)={1- 2\ep\over 1+\ep}+18\la Tr_{q}
\beta(q)^{2}{\ep(2+\ep)\over (1+\ep)^{2}}\,.
\ee
Here we indicated the dependence of $\tilde{\Gamma}$ on $\ep$ explicitely. We
note $Tr_{q} \beta(q)^{2}=\Sigma_{1}(0)$ and \Ref{sig1a} so that we obtain
finally
\be\label{gatapprox1}
\tilde{\Gamma}(\ep)={1-2\ep\over 1+\ep}+{6\ep^{2}(2+\ep)\over
  (1+\ep)^{2}} \,.
\ee
This is a positive quantity of order one (note that in between $T_{-}$ and
$T_{+}$, using \Ref{T+-} and \Ref{rangeM}, we have $\frac12\le
\ep\le\frac14$).  

Now we consider the gap equation in the extremum \Ref{gaptwo} in the leading
nontrivial approximation and obtain
\bea\label{Mhiq}
M^{2}&=&2\tilde{\Gamma}(\ep)\left(m^{2}-3\la\left({T^{2}\over 12}-{MT\over
      4\pi}+{m^{2}\over 8\pi^{2}}+{M^{2}\over 16\pi^{2}}\left(\log{(4\pi
          T)^{2}\over 2m^{2}}-2\gamma\right)\right) \nn \right. \\  
&& \left.    +6\la^{2}\gamma T^{2}
\right)\,.
\eea
This equation contains additional entries as compared to \Ref{Mquadr}.  Due to
the dependence on the mass via $\ep$ it is not longer a quadratic but a higher
order algebraic equation. Nevertheless, it is useful to rewrite it in analogy
to \Ref{M+-}. We obtain
\be\label{M+-korr}
M_{_{{Min\atop Max}}}={3\la T\over 4\pi \delta_{3}}{\tilde{\Gamma}(\epsilon)\over
  \delta_{3}}\pm \sqrt{{\tilde{\Gamma}(\epsilon)\over
  \delta_{3}}}
\sqrt{
\left({3\la T\over 4\pi \delta_{3}}\right)^{2}{\tilde{\Gamma}(\epsilon)\over
  \delta_{3}}+{2m^{2}\delta_{1}}-{\la T^{2}\over 2}\delta_{\gamma}}  \,,
\ee
with $\delta_{\gamma}\equiv 1-24\la\gamma$. Now by the same arguments as above
we obtain instead of \Ref{T+-} for the spinodal temperatures
\bea\label{T+-corr}
 T_{-}&=&{2m\over\sqrt{\la}}\sqrt{{\delta_{1}\over\delta_{\gamma}}}=
{2m\over\sqrt{\la}}\left(1-{3\la\over 16\pi^{2}}+12\la\gamma\right),
\nn \\  
T_{+}&=&{2m\over\sqrt{\la}}\sqrt{{\delta_{1}\over \delta_{\gamma}-{9\la \tilde{\Gamma}(\epsilon)\over 8\pi^2\delta_3}}}=
{2m\over\sqrt{\la}}\left(1-{3\la\over 16\pi^{2}}+ 12\la\gamma+ {9\la\tilde{\Gamma}(\epsilon)\over 16\pi^{2}}\right)\, .
\eea
Both these temperatures depend on $\gamma$ and $T_{+}$ depends in addition
on $\tilde{\Gamma}$. Their ratio is
\be\label{ratocorr}
{T_{+}\over T_{-}}=1+{9\la\over 16\pi^{2}}\tilde{\Gamma}(\ep)\,.
\ee
It remains to get some information on $\ep$.  As it comes in from $T_{+}$, we
consider $M_{Min}$ as give by equation \Ref{M+-korr} at $T=T_{+}$. Taking into
account  that we need $M_{Min}$ in the leading order ($\sim\sqrt{\la}$) only,
we obtain the equation
\[
M={3\sqrt{\la}m\over 2\pi}\tilde{\Gamma}(\ep)
\]
(at $T_{+}$ the square root in rhs. of Eq. \Ref{M+-korr} is zero). In the same
approximation we have $\ep={3\sqrt{\la}m\over 4\pi M}$ and the equation can be
rewritten as
\be\label{eqep}
{1\over 2\ep}=\tilde{\Gamma}(\ep)\,.
\ee
This is a fourth order algebraic equation with a solution
$\ep^{*}=0.318$ (one of the other solutions is negative, two are
complex). We note $\tilde{\Gamma}(\ep^{*})=1.57$. So the gap
\Ref{ratocorr} between the two spinodal temperatures becomes larger by
about $50\%$. The other quantities characterizing the phase transition
can be calculated in a similar way at last numerically. It is clear
that they become all changed by a certain amount keeping the
qualitative features unchanged.

So we arrive at the conclusion that the higher loop graphs given by Eq.
\Ref{W2-h} in comparison to \Ref{W2-1} do not change the order of the phase
transition, making it a bit stronger first order.  As a consequence, they do
not change the problem with the expansion parameter discussed at the end of
the preceding section.
  
\section{Discussion} \label{sec5}
%
Let us first discuss the main features of our analysis and the results
obtained. As it has been realized many years ago
\cite{DolanJackiw,Linde:1979px} in field theory the usual perturbation theory
in coupling constant is not applicable in the infrared region and its series
must be summed up to obtain more reliable results. A powerful method of
resummation is the second Legendre transform giving the possibility to express
the perturbative expansion in terms of two particle irreducible diagrams.
Therefore it accounts for the infrared divergencies coming from propagator
insertions.  In the present paper it has been applied to investigate the
temperature phase transition in the scalar $\phi^4(x)$ theory with spontaneous
symmetry breaking.

As an important step, the SD equation in the minimum of the free energy
functional was used. This considerably simplifies computations making them
more transparent and tractable. In order to analyze the phase transition we
adopted the ansatz \Ref{Ansatz} for the full propagator which is natural at
high temperature.  As we have seen in Sect. \ref{sec3}, all daisy and super
daisy diagrams can be summed up by taking the `eight `-diagram in $W_{2}$,
\Ref{W1}. In fact, this resummation renders the free energy in the extremum
real solving the problem of eliminating the instability connected with the
imaginary part which constitutes a necessary consistency condition. The type
of the phase transition was found to be of first order.  At the same time, as
it also follows from the analysis in Sec. \ref{sec3}, near the phase
transition the effective expansion parameter became of order one making the
contribution of subsequent graphs questionable.

To go beyond the super daisy approximation we have included into consideration
an infinite set of diagrams in the 2PI functional \Ref{W2-h} and, as a
consequence, the corresponding momentum dependent diagrams in the SD equation
\Ref{SDyone}.  Their choice is motivated by the possibility to handle them.
Besides, in an extension to the O(N)-model these graphs would constitute the
first non-leading order for large $N$. In Sec. \ref{sec4} we calculated their
contribution to the SD equation in the extremum of the free energy. Within
this approximation they turned out not to change things essentially. Their
role reduces to a redefinition of the parameters of the phase transition by a
finite factor of order of $3/2$ making the transition a bit stronger first
order. As a consequence, the estimation of the effective expansion parameter
given at the end of Sec. \ref{sec3} on the super daisy resummation level,
remains valid on the higher loop resummation of Sec. \ref{sec4} too. We come
to the conclusion that even this resummation is still insufficient to get a
reliable result from perturbation theory near the phase transition.

We believe, anyway, that the technical tools developed here, mainly the
joining of the SD equation with the necessary condition for an extremum will
find applications in gauge theories. There, as is known, the free energy is
independent of the gauge fixing parameter only in its minima. Therefore,
additional vertex resummations, such as for instance done in Ref.
\cite{Buchmuller:1995xm} in order to obtain a gauge invariant result, may be
avoided eventually. The realization of these ideas in gauge theories is a
problem left for the future.
\section*{Acknowledgment}

VS thanks the Saxonian State Ministry for Science and Arts for support
under the grant 4-7531.50-04-0361-00/12 and the University of Leipzig
for kind hospitality.

\section*{Appendix}\label{SecA1}
The function $V_{1}(M)\equiv\frac12\Tr\ln\beta$ appearing in \Ref{W1a} reads
\be\label{A1} V_{1}(M)= - \frac12 \Tr \ln \Delta 
=T\sum_{l=-\infty}^{\infty}\int{\d \vec{k}\over (2\pi)^{3}} \ln
\left((2\pi lT)^{2}+\vec{k}^{2}+M^{2}\right)\,.
\ee
Removing the \uv divergencies  the explicit expression becomes
\be\label{V1}
V_{1}(M)={1\over 64\pi^{2}}\left\{4m^{2}M^{2}+ M^{4}\left[\ln{M^{2}\over
      2m^{2}} -\frac32\right] \right\}
-{M^{2}T^{2}\over2\pi^{2}} \, S_{2}\left({M\over T}\right)\,,
\ee
where the last term is the temperature dependent contribution.  The function
$S_{2}$ can be represented as a fast converging sum over the modified Bessel
function $K_{2}$ and by an integral representation  as well:
\be\label{S2} S_{2}(x)=\sum_{n=1}^{\infty}\frac1{n^{2}} \ K_{2}(nx)={1\over
  3x^{2}} \ \int_{x}^{\infty}\ \d n \ {\left(n^{2}-x^{2}\right)^{3/2}\over
  \e^{n}-1} \,.  \ee
We use 
\[
{\pa V_{1}\over\pa V}_{|v=m/\sqrt{\la}}=0 \quad \mbox{and}\quad {\pa^{2}
  V_{1}\over\pa v^{2}}_{|v=m/\sqrt{\la}}=0\, 
\]
as normalization conditions at $T=0$. These conditions are chosen in
a way that the minimum of the free energy is at $T=0$ the same as on
the tree level.  It is possible to do so as the counter-terms can be
chosen at $T=0$, see for example a recent new proof
\cite{Kopper:2000qm}.  

The  expansion of $S_{2}(x)$ for small $x$ reads
\be\label{S2a} S_{2}(x)={\pi^4\over 45 x^2}-{\pi^2\over 12}+{\pi x\over
  6}+{x^2\over 32}\left(2\gamma-\frac32+2\ln\frac{x}{4\pi}\right)+O(x^3)\,,
\ee
giving rise to the high temperature expansion of $V_{1}$ which is at once the
expansion for small $M$:
\bea\label{V1a} V_{1}(M)&=&
{-\pi^{2}T^{4}\over 90}+{M^{2}T^{2}\over 24}-{M^{3}T\over 12\pi}  \\
&&+{1\over 64\pi^{2}}\left\{M^{4}\left[\ln{\left(4\pi T\right)^{2}\over
      2m^{2}}-2\gamma \right]+4m^{2}M^{2}\right\}+M^{2}T^{2} \ O\left({M\over
    T}\right)\,. \nn \eea
Here $\gamma$ is the Euler constant.

The function $\Sigma_{0}$ defined in Eq. \Ref{Sig0} can be obtained by
differentiating $V_{1}$
\be\label{abl}\Sigma_{0}=-{\pa\over\pa M^{2}}V_{1}(M)\,.
\ee
Its high temperature expansion is
\be\label{d0a}
\Sigma_{0}(M)={T^{2}\over 12}-{MT\over 4\pi}+{1\over
  16\pi^{2}}\left\{M^{2}\left(\ln{\left(4\pi T\right)^{2}\over
      2m^{2}}-2\gamma\right)+2m^{2}\right\}+MT \ O\left({M\over T}\right)\,.
\ee
Similar we obtain for $\Sigma_{1}(p)$ \Ref{Sig1} at $p=0$
\be\label{s1}\Sigma_{1}(0)=-{\pa\over\pa M^{2}}\Sigma_{0} \,
\ee
with the  high temperature expansion
\be\label{sig1a}
\Sigma_{1}(0)={T\over 4\pi M}+\ O\left(1\right)\,.
\ee

\bibliographystyle{unsrt}
\bibliography{Beyond}
\end{document}